\documentclass{article}
\usepackage{ijcai19}
\usepackage{nicefrac}
\usepackage{xcolor}
\usepackage{amsmath}
\usepackage{amsthm}
\usepackage{bbm}
\usepackage{amsfonts}
\usepackage[english]{babel}
\usepackage{graphicx}
\usepackage{url}

\newtheorem{theorem}{Theorem}
\newtheorem{observation}{Observation}

\newtheorem{corollary}{Corollary}
\newtheorem{algorithm}{Algorithm}
\newtheorem{lemma}{Lemma}

\theoremstyle{definition}

\newtheorem{remark}{Remark}
\newtheorem{definition}{Definition}

\newcommand{\mypara}[1]{\smallskip\noindent\textbf{#1.}}

\newcommand{\qqed}{\hfill$\square$}

\newcommand{\calR}{\mathcal{R}}

\newcommand{\iti}{{\it i}}
\newcommand{\itii}{{\it ii}}
\newcommand{\itiii}{{\it iii}}
\newcommand{\itiv}{{\it iv}}

\title{Sybil-Resilient Reality-Aware Social Choice}

\author{
 Gal Shahaf ֿֿֿֿ\\
 Weizmann Institute of Science \\
gal.shahaf@weizmann.ac.il
\and\\
Ehud Shapiro \\
Weizmann Institute of Science \\
ehud.shapiro@weizmann.ac.il
\and\\
Nimrod Talmon \\
Ben-Gurion University \\
talmonn@bgu.ac.il
}
\author{
Gal Shahaf$^1$\and
Ehud Shapiro$^1$\and
Nimrod Talmon$^2$\\
\affiliations
$^1$Weizmann Institute of Science\\
$^2$Ben-Gurion University\\
\emails
\{gal.shahaf, ehud.shapiro\}@weizmann.ac.il,
talmonn@bgu.ac.il
}

\begin{document}

\maketitle

\begin{abstract}
Sybil attacks, in which fake or duplicate identities (\emph{sybils}) infiltrate an online community, pose a serious threat to such communities, as they might tilt community-wide decisions in their favor. While the extensive research on sybil identification may help keep the fraction of sybils in such communities low, it cannot however ensure their complete eradication. Thus, our goal is to enhance social choice theory with effective group decision mechanisms for communities with bounded sybil penetration. Inspired by Reality-Aware Social Choice~\cite{rasc}, we use the status quo as the anchor of \emph{sybil resilience},  characterized by \emph{sybil safety} -- the inability of sybils to change the status quo against the will of the genuine agents, and \emph{sybil liveness} -- the ability of the genuine agents to change the status quo against the will of the sybils.
We consider the social choice settings of deciding on a single proposal, on multiple proposals, and on updating a parameter. For each, we present social choice rules that are sybil-safe and, under certain conditions, satisfy sybil-liveness.
%
%
\end{abstract}

\section{Introduction}

Our initial premise is two-fold:
  First, even though there is a vast literature concerned with identifying fake or duplicate identities, aka \emph{sybils}, one cannot assume sybils to be perfectly identified and completely eradicated.
  Second, a single vote may tilt a majoritarian group decision and as such, sybils infiltrating a group of agents that employ egalitarian democratic group decision making literally pose an existential threat to the group.
Thus, here we address the pressing need to develop group decision making processes that can be safely used in online communities that are not sybil-free.
Indeed, the vast literature on social choice proposes many aggregation methods that, unfortunately, cannot be directly used in many online settings, in which a fraction of the electorate might consist of sybils.

%

The key concept in our approach to sybil resilience is the use of the present state of affairs, namely the status quo, or \emph{Reality}, as the anchor of sybil resilience.  We characterize  \emph{sybil resilience} by \emph{sybil safety} -- the inability of sybils to change the status quo against the will of the genuine agents, and \emph{sybil liveness} -- the ability of the genuine agents to change the status quo against the will of the sybils (formal definitions in Section~\ref{section:formal model}). 

Our goal is to ensure sybil safety without sacrificing liveness, and to achieve it we follow Reality-Aware Social Choice~\cite{rasc}, which recognizes reality (i.e., the status quo) as a distinguished and ever-present alternative.

There are various settings where sybil-resilient decision making processes are needed, corresponding to different settings of social choice. As the simplest setting, we first concentrate on the case of a single proposal (i.e., an election among two alternatives, one of which is the status quo).
For this setting, we show that requiring a \emph{sybil-resilient supermajority}, defined as a simple majority plus half the sybil penetration rate, in order to change the status quo is safe.
%
%
%
Interestingly, a sybil-resilient supermajority is similar to Byzantine failures in its tipping point:
  Below one-third sybil penetration, it assures both safety and liveness, while above one-third, it assures safety but not liveness, as sybils, while unable to force a change to the status quo, may block any change to it. 

We then consider ordinal elections for deciding among multiple alternatives, one of which is the status quo (reality). We describe an efficient Amendment Agenda that is safe and provides liveness when sybil penetration is under one third.
%
%
Finally, we consider sybil-resilience when deciding upon the value of a parameter, e.g., the interest rate or inflation rate of a sovereign currency, the conductance and solidarity of an expanding e-community~\cite{poupko2019sybil},  the votes threshold for parties in a parliamentary system, or the gas price of a cryptocurrency.
Assuming single-peakedness for this setting,
we describe a rule that, briefly put, disregards sufficiently-many extreme votes, and show it to be sybil-safe.

\mypara{Related Work}\label{section:related work}
There is a vast literature on defending against sybil attacks, see, e.g., recent surveys~\cite{sybilsurvey,viswanath2010analysis}.
That literature is usually concerned with graphs on which the genuine and sybil entities reside, and the focus is usually not on group decision making.
E.g., Douceur~\shortcite{douceur2002sybil} describes a very general model for studying sybil resilience and presents some initial negative results in this model. Many papers consider leveraging graph properties such as various centrality measures to identify suspicious nodes (see, e.g.,~\cite{cao2012aiding}). As further examples, Molavi et al.~\shortcite{molavi2013iolaus} aim to shield online ranking sites from the negative effects of sybils and Chiang et al.~\shortcite{chiang2013secure} consider sybil-resilience in the context of radio networks.

We are particularly interested in sybil-resilient group decision making.
This scenario is considered by Tran et al.~\shortcite{tran2009sybil}, but with a different goal and solution:
  While we aim to protect democratic decisions from sybil attacks, they are considering ranking online content.  
Other relevant papers are the paper of Conitzer and Yokoo~\shortcite{conitzer2010using}, concentrating on axiomatic characterizations of sybil-resilient rules in a certain formal model. In essence, Conitzer et al. show that in a model without a distinguished status quo alternative, the only voting rules which are sybil-safe, in the sense that there is no incentive for an attacker to produce sybils, is of the form ``if all vote unanimously for $c$, pick $c$, otherwise pick a winner at random''. Indeed, this negative result can also be seen as a motivation for our model of sybil-safety, which does incorporate the status quo as a distinguished alternative, as this allows for a conservative default to the status quo, rendering the negative result of Conitzer et al.\  inapplicable.
Related papers exist~\cite{wagman2008optimal,wagman2014false,waggoner2012evaluating,conitzer2010false,conitzer2008anonymity}.

We also mention the vast literature on control and bribery in elections~\cite{controlandbribery}, studying malicious entities aiming at changing elections outcomes (we also mention recent work connecting bribery to robustness measures of voting rules~\cite{bredereck2017robustness,faliszewski2017bribery}).  The model of election control assumes a given voting rule and a given electorate, and the question is whether an external agent, called the chair of the election, may change the election structure, e.g. by adding or removing candidates or votes,  to have its preferred candidate win (or lose). The model of a sybil attack is that of an external agent that cannot change the vote structure, but has control of the actual votes of a fraction of the electorate (the sybils and their creators/perpetrators). Hence, formally, a sybil attack by a fraction $\sigma\in [0,1]$ of the voters is similar to election control where the chair may add up to a fraction $\sigma$ of the voters. However, rather than studying how this specific form of control may affect existing voting rules, we design new voting rules that are resilient to sybil attacks, a notion defined below.


\section{Abstract Model}\label{section:formal model}

In this paper we use disjoint union $X = Y \uplus Z$ as a shorthand for $X = Y \cup Z$, $Y \cap Z = \emptyset$. Our model is as follows.
%
We assume a set of agents $V = H \uplus S$ which is a union of two disjoint sets, the set of genuine agents $H$ and the set of sybils $S$.
We wish to design sybil-resilient voting rules for the agents $V$.
We assume that all agents participate in every vote\footnote{A forthcoming paper explores how proxy voting/vote delegation can be employed to relax this assumption and still retain sybil resilience.}, so we overload the notation and identify the agent $v \in V$ with its vote.
%
We follow Reality-Aware Social Choice in considering decisions on a set of alternatives~$A$ that always includes the status quo $r \in A$ (reality) as a distinguished, ever-present alternative.
Given such a set $A$ and $n$ votes over it, a \emph{voting rule} returns a set of alternatives as the co-winners of the election. The specific set of allowed alternatives and the mathematical objects modeling a vote are different for each social choice setting we consider; we elaborate on these in the corresponding sections, but first we make an abstract exposition.


\subsection{Sybil Safety and Sybil Liveness}

We wish to have voting rules that are sybil safe, in the sense that they prevent sybils from changing the status quo against the will of the genuine agents.  But how is the will of the genuine agents defined?  Presumably, via en established voting rule, e.g., the majority rule when voting on a single proposal against the status quo, or some social choice function when voting on multiple alternatives.  The following definition aims to capture this intent by defining a voting rule to be safe (with respect to a base voting rule) if it elects an alternative to the status quo when applied to votes of all agents only if the base rule may elect this alternative when applied to a subset of the votes  -- the votes of the genuine agents.

\begin{definition}[Sybil Safety]\label{definition:Sybil Safety}
Consider a set of alternatives $A$ with reality $r \in A$, a set of agents $V = H \uplus S$, and let $\mathcal{R}$ and $\mathcal{R}'$ be two voting rules.
Then, the voting rule $\mathcal{R}$ is \emph{sybil safe} with respect to $\mathcal{R'}$, or \emph{safe} for short, if the following holds:
  If   $\mathcal{R}(V) \cap A \setminus \{r\} \neq \emptyset$
  , then $\mathcal{R}(V) \subseteq \mathcal{R'}(H)$.
That is, if $\mathcal{R}$ chooses some alternative $a\neq r$, then $a$ is chosen also by $\mathcal{R}'$ over the honest voters.
\end{definition}

\begin{remark}
Below we consider three social choice settings: voting on one proposal, voting on multiple alternatives, and voting on the value of a parameter.  For each setting we chose a base voting rule that is suitable for the domain, employing three criteria: (\iti) Broad recognition (\itii) Simplicity; (\itiii) Ease of attaining safety.  Specifically,  we rely on May's theorem~\cite{may1952set}, the Condorcet criterion~\cite{gehrlein1985condorcet}, and Black's theorem~\cite{black1948rationale},respectively, in choosing the base rules for the three settings.

We wish to stress the importance of simplicity: The  trust of voters in the voting process critically depends on their understanding of it.  Hence a voting rule must be easy to communicate, even at the expense of other desirable properties that can be achieved only through complications.\qqed  
\end{remark}

As sybil safety can be achieved trivially by sticking with the status quo, it must be combined with a liveness requirement -- that the genuine agents are able to change the status quo despite the sybils.

\begin{definition}[Sybil Liveness]
Consider a set of agents $V = H \uplus S$, a set of alternatives $A$, reality $r \in A$, and a voting rule~$\mathcal{R}$. 
We say that $\mathcal{R}$ satisfies \emph{sybil liveness} for $V$ and $A$, or \emph{liveness} for short, if, for any set of votes of the sybils $S$ and for any alternative $a \in A \setminus \{r\}$, there is a set of votes of the genuine agents for which $\mathcal{R}$, applied to all agents, elects $a$.
\end{definition}

Note that while safety is being defined with respect to a base rule, liveness isn't, as liveness merely makes sure that some progress can be made according to the current rule; safety than ensures any such progress is safe with respect to the base rule.

%

%
%

We use the term \emph{sybil resilience} to refer jointly to sybil safety and sybil liveness. For each of the settings we consider here, our main goal is:
  \emph{Ensure sybil resilience without being unnecessarily conservative in defending the status quo.}
The following definition captures a specific aspect of sybil resilience.

\begin{definition}[Sybil-Penetration Resilience]
A voting rule $\mathcal{R}$ is \emph{resilient to the penetration of up to $\sigma$ sybils} with respect to a base voting rule $\mathcal{R'}$, if it ensures sybil safety with respect to $\mathcal{R'}$ and sybil liveness for every set of agents $V = H \uplus S$, provided the sybil penetration rate is below $\sigma$, namely $\frac{|S|}{|V|} < \sigma$.
\end{definition}

\begin{remark}
How to estimate the sybil penetration $\sigma$ is an important question. While in some cases there might be other techniques available, usually it is natural to assume that by sampling a voter one can estimate the probability that the voter is genuine or fake (e.g., looking at her Facebook profile). Thus, the main general technique we suggest is to sample voters uniformly at random and, given the sampling results, estimate $\sigma$. Note that using such sampling it is then possible to compute, for a given value $p$, a value $z$, such that the probability that $\sigma$ is greater than $z$ is at most $p$. Alternatively, one can compute the mean $m$ of the sample and take an $\epsilon$ margin of safety, i.e., use $m + \epsilon$ as the estimate for $\sigma$.
\end{remark}

\section{Sybil-Resilience for One Proposal}\label{section:singleproposal}

We begin our investigation with  \emph{yes/no} decisions on a single proposal $p$, where a \emph{yes} vote favors $p$ and a \emph{no} vote favors the status quo (e.g., Brexit vs. Remain).
So, formally, the set of alternatives is $A = \{p, r\}$, and each vote $v \in V$ is either $v = p$ or $v = r$.
For this setting, it is natural to use supermajority as the basis for a sybil-safe decision rule, and to use simple majority as the base decision rule against which sybil-safely is measured.

\begin{definition}[$\delta$-Supermajority]
In a decision on a proposal~$p$ against the status quo $r$, the proposal $p$ is said to \emph{win by a $\delta$-supermajority}, $\delta \in [0,\nicefrac{1}{2}]$, if  more than $\nicefrac{1}{2}+\delta$ of the agents prefer $p$ over $r$ (i.e., vote for $p$).
The proposal \emph{wins by a simple majority} if it wins by a $0$-supermajority.
\end{definition}

\begin{definition}[Reality-Aware $\delta$-Supermajority Rule]\label{definition: Reality-aware delta-Supermajority Rule}
When deciding on a single proposal $p$ against the status quo $r$, the \emph{reality-aware $\delta$-supermajority rule} elects $p$ if it is preferred over $r$ by a $\delta$-supermajority, else it elects the status quo $r$.
The reality-aware $0$-supermajority rule is referred to as the \emph{majority rule}. 
\end{definition}

\begin{remark}
Notice that $\delta$-supermajority rule with $\delta > 0$ follows Reality-Aware Social Choice in favoring the status quo.\qqed
\end{remark}

Requiring $\delta = \nicefrac{1}{2} - \epsilon$, $\epsilon < \frac{1}{|V|}$, would render the reality-aware $\delta$-supermajority rule sybil safe, as it would elect the proposal only if all agents are in favor of it; it would, however, be unnecessarily conservative. Next we characterize the minimal~$\delta$ needed for safety.
%

\begin{lemma}[Safety of Supermajority]\label{lemma: all vote}
Let $V = H \uplus S$ be the set of agents, $\sigma = \frac{|S|}{|V|}$, and let $p$ and $r$ be a proposal and the status quo.
Then, if $p$ is preferred over $r$ by a $\nicefrac{\sigma}{2}$-supermajority of all agents, then $p$ is preferred over~$r$ by a majority of the genuine agents.
\end{lemma}

\begin{proof}
Consider the equation:
$$\nicefrac{1}{2} + \delta = 
	\frac{\sigma + 
		\nicefrac{1}{2} \cdot (1 - \sigma)}
        {\sigma+(1 - \sigma)}\ ,$$ 
with the left side of the equation being the  $\delta$-supermajority required for the majority of the genuine agents to vote for the proposal, assuming all sybils also vote for it, and with the right side being the sybils ($\sigma$) and the  majority ($\nicefrac{1}{2}$) of the genuine agents $(1-\sigma)$, divided by the total agents, namely the sybils ($\sigma$) and the genuine agents ($1-\sigma$).
Solving for $\delta$ gives $\delta = \nicefrac{\sigma}{2}$. 
\end{proof}

\begin{remark}
The value $\nicefrac{\sigma}{2}$ above is tight, as any value strictly smaller than $\nicefrac{\sigma}{2}$ would not be safe. To see this, assume that all sybils, as well as slightly less than half of the genuine agents, vote in favor of the proposal $p$.
\end{remark}

\begin{theorem}[Safety of Reality-Aware Supermajority Rule]\label{theorem: all vote}
Let $V = H \uplus S$ be the set of agents and $\sigma = \frac{|S|}{|V|}$. Then, the reality-aware $\nicefrac{\sigma}{2}$-supermajority rule is safe with respect to the majority rule.
\end{theorem}
\begin{proof}
Follows Lemma~\ref{lemma: all vote} and Definitions \ref{definition:Sybil Safety} and \ref{definition: Reality-aware delta-Supermajority Rule}.
\end{proof}

Next, we offer a measure for the conservatism of a supermajority rule, by investigating the situations in which the genuine agents can indeed change the status quo. 

\begin{definition}[Supermajority Conservatism]
Let $V = H \uplus S$ be the set of agents and let $\mathcal{R}$ be a reality-aware supermajority voting rule. The \emph{conservatism} $\rho$ of $\mathcal{R}$ is defined as the supermajority among the genuine agents needed in order to change the status quo, according to~$\mathcal{R}$, assuming all sybils vote in favor of the status quo.
\end{definition}


\begin{observation}\label{observation:one}
The conservatism of the reality-aware $\delta$-supermajority rule, given a sybil penetration rate $\sigma$, is
  $$\rho = \frac{\nicefrac{1}{2}+\delta}{1 - \sigma}-\nicefrac{1}{2.}$$  
\end{observation}

\begin{proof}
Let $V = H \uplus S$ be the set of agents, $\sigma = \frac{|S|}{|V|}$, and consider the reality-aware $\delta$-supermajority rule where we have $\sigma n$ sybils, all voting in favor of the status quo, and $(1 - \sigma) n$ genuine agents. Then, for a $\rho$-supermajority among the genuine agents, which is exactly $(1 - \sigma) n (\nicefrac{1}{2} + \rho)$ genuine agents voting for the proposal to change the status quo, they shall constitute at least a $(\frac{1}{2} + \frac{\sigma}{2})$-fraction of the full electorate, which contains $n$ agents. Thus, solving the equation
$$
(1 - \sigma) n \left(\frac{1}{2} + \rho\right) = \left(\frac{1}{2} + \delta\right) n
$$
for $\rho$ gives the result.
\end{proof}

\begin{figure}
    \centering
    \includegraphics[scale=0.4]{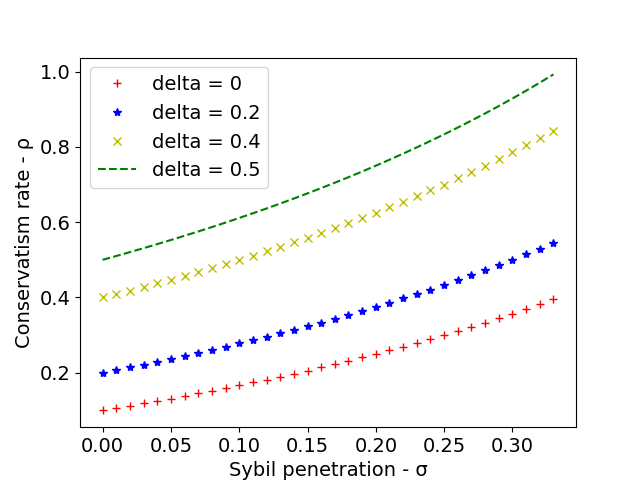}
    \caption{Conservatism rate $\rho$ as a function of $\sigma$ and $\delta$.}
    \label{figure:one}
\end{figure}

Figure~\ref{figure:one} depicts the conservatism rate $\rho$ as given by the formula in Observation~\ref{observation:one}.

\begin{remark}
Of particular interest is the special case $\delta=\nicefrac{\sigma}{2}$, which, following Observation~\ref{observation:one}, implies a conservatism of $\rho = \frac{\sigma}{1-\sigma}$. Notice that: (\iti) If there are no sybils, then $\rho = 0$, which corresponds to a simple majority.
(\itii) On the other extreme, if a $\nicefrac{1}{3}$-fraction of the agents are sybils, then  $\rho = \nicefrac{1}{2}$, meaning that
the proposal cannot be chosen even if all genuine agents are unanimously for it, violating liveness.  The same is true of course if more than $\nicefrac{1}{3}$-fraction of the agents are sybils.
(\itiii) With single-digit sybil-penetration, i.e. $\sigma < 10\%$, the above gives $\rho < \nicefrac{1}{9}$, hence the supermajority needed among genuine agents would be under $61.2\%$, which is quite reasonable.
\end{remark}

\begin{corollary}[Supermajority Liveness]
  Let $V = H~\uplus~S$ be the set of agents and let $\sigma = \frac{|S|}{|V|}$. If all genuine agents vote, then the reality-aware $\nicefrac{\sigma}{2}$-supermajority rule satisfies sybil liveness if and only if $\sigma < \nicefrac{1}{3}$.
\end{corollary}

\begin{proof}
Following Observation~\ref{observation:one}, we have that $\rho = \frac{\sigma}{1 - \sigma}$. Solving $\frac{\sigma}{1 - \sigma} < \nicefrac{1}{2}$ for $\sigma$, corresponding to almost $\nicefrac{1}{2}$-supermajority (unanimity among the genuine agents), gives $\sigma < \nicefrac{1}{3}$.
\end{proof}

\begin{corollary}[Supermajority Resilience]\label{corollary:two}
 The $\nicefrac{\sigma}{2}$-supermajority rule is resilient to a penetration of up to $\sigma = \nicefrac{1}{3}$ sybils.
\end{corollary}

Hence, we refer to the reality-aware $\nicefrac{\sigma}{2}$-supermajority, with $\sigma < \nicefrac{1}{3}$, as \emph{sybil-resilient supermajority}.

\begin{remark}
As with Byzantine failures, a sybil penetration of $\sigma = \nicefrac{1}{3}$ is an inflection point wrt.\ sybil-resilience of $\nicefrac{\sigma}{2}$-supermajority:
  Up to $\nicefrac{1}{3}$ sybils,  a simple majority among the genuine agents can defend the status quo, i.e., veto a change to it, and a sufficiently large supermajority of the genuine agents may change the status quo. So the sybils can neither enforce a change nor veto one, if the genuine agents are sufficiently determined and united.
From $\nicefrac{1}{3}$ sybils and above, however, the sybils have a veto right:
  If the sybils unanimously object to a change, then no majority of the genuine agents can effect it.
\end{remark}

\section{Sybil-Resilient Ordinal Elections}\label{section:multipleproposals}

We assume the ordinal model of elections, thus each vote is a ranking over the set of alternatives $A$ that includes the status quo $r \in A$.  
Formally, denoting the set of all rankings over a set of alternatives $A$ by $L(A)$, we define a \emph{voting rule} to be a function $\mathcal{R}: L(A)^n \rightarrow 2^A$ that takes $n$ ordinal votes over $A$ and returns a set of tied elected alternatives. If a singleton is elected from $A$, then it is referred to as the \emph{winner} of $\mathcal{R}$ for the election. Otherwise, each of the alternatives returned from $\mathcal{R}$ is referred to as a co-winner of the election. Notice that, for technical reasons, we do not consider tie breaking.
Our approach to sybil-resilience for this setting is to adapt the Condorcet principle. We need the following definition first.

\begin{definition}[Reality-Viable Alternatives]\label{definition:delta-reality-viable}
Let $V$ be a set of agents, $A$ a set of alternatives with $r \in A$ the reality, and let $\delta \in [0,\nicefrac{1}{2}]$.
An alternative $a \in A$ is \emph{$\delta$-reality-viable} (\emph{$\delta$-viable} for short) if $a$ beats $r$ by a $\delta$-supermajority; i.e., if at least a $\nicefrac{1}{2}+\delta$-fraction of the voters (weakly) prefer $a$ over $r$. We denote the set of $\delta$-reality-viable alternatives by $A_r^\delta$. If $\delta = 0$, $\delta$ can be omitted and the set of reality-viable alternatives $A_r$ is defined via a simple majority. 
\end{definition}

The following definition presents two variants\footnote{%
  Other variants, such as using any tournament solution, are also possible but not explored here.} of a Reality-aware Condorcet voting rule, which will serve as the base rules against which we will measure sybil resilience.

\begin{definition}[Reality-Aware Condorcet Rule]\label{definition:condorcet conservative rule}
Let $A$ be a set of alternatives with $r \in A$. If $A_r$ has a Condorcet winner, then elect it. Else, either:

\begin{enumerate}
    \item (Conservative rule) elect~$r$.
    \item (Permissive rule) elect all of $A_r$ as co-winners.
\end{enumerate}

\end{definition}

We will adapt these Reality-aware Condorcet voting rules to be sybil-resilient by replacing simple majority by a $\delta$-supermajority.  But first we need to adapt the notion of a Condorcet winner to use $\delta$-supermajorities.


\begin{definition}[$\delta$-Supermajority Condorcet winner]\label{definition:condorcet supermajority winner}
Let $A$ be a set of alternatives and let $\delta \in [0,\nicefrac{1}{2}]$.  
An alternative $a \in A$ is a \emph{$\delta$-supermajority Condorcet winner} if $a$ is preferred over any $a' \in A$, $a \ne a'$, by a $\delta$-supermajority.
\end{definition}

Our approach extends the Condorcet principle as it is adapted to Reality-Aware Social Choice by employing supermajorities. Specifically, next we discuss several reality-aware Condorcet criteria adapted to our setting via $\delta$-supermajorities; voting rules that adhere to these criteria follow in a straightforward way.

\begin{definition}[Reality-Aware $\delta$-Supermajority Condorcet criterion]\label{definition:Conservative Reality-aware delta-Supermajority Condorcet criterion}
Let $A$ be a set of alternatives with $r \in A$, and let $\delta \in [0,\nicefrac{1}{2}]$. If $A_r^\delta$ has a $\delta$-supermajority Condorcet winner then elect it. Else, either:

\begin{enumerate}
    \item (Conservative criterion) elect~$r$.
    \item (Permissive criterion) elect all of $A_r^\delta$ as co-winners.
\end{enumerate}

\end{definition}

%
Notice that all variants of the Reality-Aware $0$-supermajority Condorcet criteria (i.e., where $\delta = 0$) are identical to the variants of the Reality-Aware Condorcet criteria (Definition \ref{definition:condorcet conservative rule}).  

%

The next theorem characterizes the minimal~$\delta$ for which voting rules satisfying the above criteria are safe with respect to the base Condorcet rules defined earlier. 
Notice that Definition~\ref{definition:Conservative Reality-aware delta-Supermajority Condorcet criterion} is concerned with sybil-resilience and incorporate $\delta$-supermajorities. In contrast, Definition~\ref{definition:condorcet conservative rule} employs simple majorities.

\begin{theorem}\label{theorem: Condorcet all vote}
Let $V = H \uplus S$ be the set of agents and let $\sigma = \frac{|S|}{|V|}$. Then, a voting rule satisfying Reality-Aware $\nicefrac{\sigma}{2}$-supermajority Condorcet criterion (Definition \ref{definition:Conservative Reality-aware delta-Supermajority Condorcet criterion}) is safe with respect to the Conservative Reality-Aware Condorcet rule (Definition \ref{definition:condorcet conservative rule}).
\end{theorem}

\begin{proof}
Let $A$ be the set of alternatives with the reality being $r \in A$, let $\mathcal{R}$ be a rule satisfying the Reality-Aware $\nicefrac{\sigma}{2}$-Supermajority Condorcet Criterion, and let $c$ be its winner in a given election. If $c = r$, then we are done as electing the status quo is always safe. Else, if $c \neq r$, then $c$ wins over each alternative in $A^{\nicefrac{\sigma}{2}}_r$ by a $\nicefrac{\sigma}{2}$-Supermajority. Hence, by Lemma \ref{lemma: all vote}, $c$ wins over all these alternatives by a simple majority among the genuine agents; thus, if there is a Condorcet winner among the genuine agents then it must be $c$. So, the Reality-Aware Condorcet Rule would elect either $c$ or $r$.
\end{proof}

The following Observation \ref{observation:two} and Corollary \ref{corollary:three} follow a reasoning similar  to Observation~\ref{observation:one} and Corollary~\ref{corollary:two}. 

\begin{observation}\label{observation:two}
The conservatism of a Reality-Aware $\delta$-Supermajority Condorcet consistent rule, given a penetration rate $\sigma$ of sybils, is $\rho = \frac{\nicefrac{1}{2}+\delta}{1 - \sigma}-\nicefrac{1}{2}$.  
\end{observation}

\begin{corollary}\label{corollary:three}
 A Reality-Aware $\nicefrac{\sigma}{2}$-Supermajority Condorcet rule is resilient to a penetration of up to $\sigma = \nicefrac{1}{3}$ sybils.
\end{corollary}

\begin{remark}
While here we consider only linear orders, it is possible to extend the analysis to accommodate partial orders, including weak rankings and 1-Approval ballots (where each voter declares her most preferred alternative).
\end{remark}

\mypara{An Efficient Sybil-Resilient Amendment Agenda}\label{section:SRAA}
For concreteness and for practical applications, we complement the discussion with an efficient algorithmic realization of the supermajority Condorcet criteria defined above. Our realization is based on Llull's Amendment Agenda~(1299, cf.~\cite{mclean1990borda}): 
  Arrange all alternatives in some order, vote the first against the second, the winner of the two against the third, and so on,  then elect the final winner.
The Amendment Agenda is Condorcet consistent.
We make four enhancements to this Agenda: (\iti) We consider only $\delta$-reality-viable alternatives; (\itii) we start with the reality $r$; (\itiii) we employ sybil-resilient supermajorities; and (\itiv) at the end we check for a Condorcet top-cycle, and resort to reality if one is detected.

\begin{algorithm}[Conservative $\delta$-Supermajority Amendment Agenda]\label{algorithm:SRAA}
Let $A$ be the set of alternatives with $r \in A$ and let  $\delta \in [0,\nicefrac{1}{2}]$.  If $A_r^\delta = \emptyset$, elect $r$. Else, perform an Amendment Agenda vote on $A_r^\delta$ starting with~$r$ and employing $\delta$-supermajorities, and let $w \in A_r^\delta$ be the winner. Then, vote $w$ against all members of $A_r^\delta$ not previously voted against $w$, if any. If $w$ wins all these votes by a $\delta$-supermajority then elect $w$.
Else elect~$r$.
\end{algorithm}

\begin{theorem}\label{theorem:SRAA}
Algorithm \ref{algorithm:SRAA} satisfies the Conservative Reality-Aware $\delta$-Supermajority Condorcet criterion.   
\end{theorem}

\begin{proof}
We do a case analysis.
First, if $A_r^\delta = \emptyset$, then the Agenda elects $r$; this is what the axiom dictates.
Otherwise, i.e., if $A_r^\delta \neq \emptyset$ then there are two cases to consider:
  First, if there is a $\delta$-supermajority Condorcet winner $c$, then $c$ will be the final winner as it will not be eliminated during the agenda and will also beat all alternatives in the check at the end of the Agenda, thus $c$ will be elected as dictated by the criterion. Otherwise, the final winner $w$ of the Agenda is not a $\delta$-supermajority Condorcet winner, and thus there will be at least one alternative $w'$ for which $w$ will not win by a $\delta$-supermajority, thus $r$ will be elected, again as dictated by the criterion.
\end{proof}

\begin{remark}
An Amendment Agenda corresponding to the permissive $\delta$-supermajority Condorcet rule can be obtained by revising the final ``Else'' clause to be   ``Else arbitrarily elect a member of $A_r^\delta$.''   The proof is similar.  Many other sybil-resilient tournament solutions can be obtained by revising the ``Else'' clause appropriately.
\end{remark}

\begin{remark}
Observe that our results for a single proposal (Section~\ref{section:singleproposal}) carry over to the sybil-resilient Condorcet criteria: For one proposal, all variants of the Reality-Aware $\delta$-Supermajority Condorcet rule as well as our sybil-resilient amendment agenda boil down to the one-proposal $\delta$-supermajority rule.
\end{remark}

\section{Sybil-Resilient Parameter Update}

We consider sybil-resilience when deciding upon the value of a parameter, e.g., the target inflation rate, the interest rate of a sovereign currency, the gas price of a cryptocurrency, the conductance and solidarity of an expanding e-community~\cite{poupko2019sybil}, or the votes threshold for a party in a parliamentary system.  In all these examples, we may assume that each voter has a preferred value for the parameter (an \emph{ideal point}), and the closer the elected value to the ideal point, the happier the voter.

We model such settings by considering a one-dimensional single-peaked domain; specifically, we assume that the parameter can take real-valued numbers, that each voter has a single ideal point $v \in \mathbb{R}$ which she declares as her vote, and single-peakedness then means that a voter with ideal point $v$ prefers some $y$ to $z$ if $v \leq y < z$ or if $z < y \le v$.
We stress that, contrary to the setting of Section~\ref{section:multipleproposals}, here voters declare only their ideal points and not their rankings.
The assumption of single-peakedness then allows us to devise sybil-resilient voting rules for this setting despite the fact the the domain of alternatives is infinite.

Black's Median Voter Theorem~\cite{black1948rationale} states that within this model, the ideal point of the median voter is the unique unbeaten point and the Condorcet winner.
Consider electing the value of the parameter \emph{de novo}. How can it be made sybil-safe if even a single sybil may affect the identity of the median voter and, furthermore, it cannot be determined whether such a sybil has tilted the median to be higher or lower?
Lacking an answer and being inspired by Reality-Aware Social Choice we, therefore, forgo \emph{de novo} parameter election and consider the problem of \emph{parameter update}:
  Given the current value of a parameter, how can its value be updated in a sybil-resilient way?
Formally, we aim at designing a \emph{parameter update rule} $\calR$, which is a function that takes the current parameter value $r$ and a set of $n$ votes and returns a new value for the parameter (all values in $\mathbb{R}$).

As before, we are interested in sybil-safety, which abstractly means that the current value of the parameter shall change only if the genuine voters wish so. Following Black's Median Voter Theorem, we wish to use the median rule as the base rule. But, to overcome the limitation of the median being well-defined only for an odd number of voters, we employ reality as follows.

\begin{definition}[Reality-Aware Median]
Let $r$ be the current value of the parameter and $V$ be the set of votes.
The \emph{reality-aware median} $v^*$ of $V$ is the median of $V$ if $|V|$ is odd and the median of $V \cup \{r\}$ otherwise.
\end{definition}

\begin{remark}
The effect of the definition for an even number of $2k$ ordered votes is as follows: If $v_k \le v_{k+1} \le r$, then the reality-aware median is $v^* := v_{k+1}$;
if $r \le  v_k \le v_{k+1}$, then $v^* := v_k$; and if $v_k \le r \le v_{k+1}$, then $v^* := r$.
This means that the present value $r$ breaks ties in its favour, and in particular if half the voters wish to increase the parameter and half to decrease it, the present value of the parameter remains, as it should.\qqed
\end{remark}

We use the reality-aware median to define the base rule against which we will measure sybil safety.

\begin{definition}[Reality-Aware Median Base Rule]
Let $r$ be the current value of the parameter, $V$ be the set of votes, and $v^*$ the reality-aware median of the voters.
If $r \le v*$, then the \emph{Reality-Aware Median Base Rule} returns the set $\{v \in V : r \leq v \leq v^*\}$, and if $v^* \leq r$, then it returns the set $\{v \in V : v^* \leq v \leq r\}$.
\end{definition}

Namely, a parameter update rule is safe wrt.\ the reality-aware median base rule if it does not change the value of the parameter further than the reality-aware median $v^*$ of the genuine agents or in an opposite direction to it.  Indeed, the degenerate rule that never changes the parameter value is safe; liveness then considers the ability of the genuine agents to change the value of the parameter in their preferred direction despite the sybils.
Notice how, informally speaking, the single-peakedness assumption allows speaking of ``directions'' and not be confined to Condorcet winners as in Section~\ref{section:multipleproposals}.

\mypara{A Simple Update Rule}
We first present a simple update rule, which only considers the ``directions''.

\begin{definition}[Simple Update Rule]
Let $r$ be the current value of the parameter, $V$ be the set of votes and $\sigma \in [0,1]$. If there is a $\frac{\sigma}{2}$-supermajority of ideal points larger (smaller) than $r$, then select the smallest ideal point larger than $r$ (respectively, the largest ideal point  smaller than $r$); otherwise, select~$r$. 
\end{definition}

\begin{remark}
The Simple Update Rule can be seen as a $\sigma/2$-supermajority rule for the case of two proposals against the status quo, namely $p^-$ and $p^+$ against $r$, where it is assumed that a voter voting for $p^-$ prefers $r$ over $p^+$ and a voter voting for $p^+$ prefers $r$ over $p^-$.
\end{remark}

\begin{observation}
  The Simple Update Rule is sybil-safe and satisfies liveness whenever $\sigma < \nicefrac{1}{3}$.
\end{observation}

\mypara{A Least-Conservative Update Rule}
The simple update rule satisfies liveness. However, it is quite conservative in that it moves in ``baby steps''. It is natural to seek a parameter update rule that not only updates the parameter in the right direction, but also pushes its value as far as sybil-safety permits.

\begin{definition}
Let $\calR$ and $\calR'$ be two parameter update rules. Then, $\calR$ is \emph{less conservative} than $\calR'$ if for every set of votes $V$ and current parameter value $r$, the updated values obtained by these rules satisfy either $r \leq \calR'(V,r) \leq \calR(V,r)$ or $ \calR(V,r) \leq \calR'(V,r) \leq  r$.
\end{definition}

Our approach to achieve lesser conservatism is as follows:
  If the median of the ideal points of all agents is above the current value, we make the worst-case assumption that all sybils wish to extremely increase the parameter value; we therefore remove the top $\sigma$ values and recompute the new median.
  If the recomputed median is still above the current parameter value, then it is safe to elect it; otherwise, we revert to the status quo.  Suppressing these extreme votes can be justified to voters by saying that, in the worst case, all these votes could be by sybils and hence, to be on the safe side, we must ignore them.

\begin{definition}[Reality-Aware Median with Outer-$\sigma$ Suppression]
Let $r$ be the current parameter value,   $V$ be a set of voters and  $\sigma \in [0,1]$.
Then the set $V^{-\sigma}$ is obtained by removing from $V$ its top $\sigma$-fraction, the set $V_{-\sigma}$ is obtained by removing from $V$ its bottom $\sigma$-fraction, $v^{-\sigma}$ is the reality-aware median of $V^{-\sigma}$, and $v_{-\sigma}$ is the reality-aware median of $V_{-\sigma}$.
\end{definition}

\begin{definition}[Suppress Outer-$\sigma$ Parameter Update Rule]
Let $r$ be the current parameter value and $V$ be the set of votes with reality-aware median $v^*$. 
Then, the \emph{Suppress Outer-$\sigma$ parameter update rule} is defined as follows:
  If $r < v^{-\sigma} \le v^*$, then update the parameter to be $v^{-\sigma}$;
  if $v^* \le v_{-\sigma} < r$, then update the parameter to be $v_{-\sigma}$;
  otherwise keep the current parameter value~$r$.
\end{definition}

\begin{theorem}[Sybil Resilience of the Suppress Outer-$\sigma$ Rule]
The Suppress Outer-$\sigma$ Parameter Update Rule is resilient up to $\sigma < \nicefrac{1}{3}$ sybil penetration.
\end{theorem}

\begin{proof}
Assume a current value $r$ and a set of agents $V = H \uplus S$.  There are three possible outcomes to the Suppress Outer-$\sigma$ rule: $v^{-\sigma}$, $v_{-\sigma}$, and $r$.

For sybil-safety,
consider the first outcome $v^{-\sigma}$. As in this case $r < v^{-\sigma}$, what is left to show is that $v^{-\sigma} \le v^*$.
We consider two sub-cases:
(\iti)~There are no sybils left in $V^{-\sigma}$ greater than $v^*$.  In this case the difference between $V^{-\sigma}$ and $H$ are top genuine votes that are in $H$ but eliminated from $V^{-\sigma}$, if any, and sybils votes smaller than $v^*$ in $V^{-\sigma}\setminus H$, if any. Hence $v^{-\sigma} \le v^*$ as required.
(\itii)~There are sybils in $V^{-\sigma}$ greater than $v^*$.  In this case, there must be at least as many top genuine identities eliminated from $V^{-\sigma}$, since $\sigma$ is a bound on the number of sybils. Now let us swap the type (genuine/sybil), but not the vote, of such sybil and genuine votes, so no sybil votes greater than $v^*$ are left in $V^{-\sigma}$.  Doing so would not affect $v^{-\sigma}$, as it is ``type-blind'', and would not affect $v^*$ since all pairs of type-swapped votes are greater than $v^*$.
And we are now in sub-case (\iti) which has been proved.
The safety of the second outcome $v_{-\sigma}$ is proved symmetrically, while the third outcome, $r$, is safe by definition.
For sybil-liveness,
assume that $\sigma < \frac{1}{3}$ and that all genuine agents vote for a certain value $q$ above $r$.  Since $|H| > \frac{2}{3}|V|$ and $V^{-\sigma}$ eliminates at most $\sigma\cdot |V| < \frac{1}{3}|V|$ of the genuine voters, the genuine votes will be a majority in $V^{-\sigma}$, and hence its
median $v^{-\sigma}$ will be larger than $r$, specifically $q$, resulting in the update to $v^{-\sigma}$.  The symmetric argument applies if all genuine votes are some $q$ below $r$.  
\end{proof}

Next we argue that indeed the rule defined above is the least-conservative update rule.

\begin{remark}
In particular, if $V=v_1\leq v_2\leq \ldots \leq v_n$, the reality aware median may always assume that $n$ is odd (if even, it just adds reality to $V$). Hence, $v_*:=v_{\lceil n/2 \rceil}$, $v^{-\sigma} :=v_{\lceil n/2 + \sigma n /2 \rceil}$, and $v^{-\sigma} :=v_{\lceil n/2 - \sigma n /2 \rceil}$. Under this notation, it follows that any voting rule that elect $x$, $v^{-\sigma} < x$, when $r<v^*$ is not safe, because whenever the honest voters are $v_1\leq v_2\leq ...\leq v_{n(1-\sigma)}$, then $v^* = v_{\lceil n/2 - \sigma n /2 \rceil} = v^{-\sigma}$, and thus $r \leq v^* < x$.
\end{remark}

\section{Discussion}

While a single fake agent may tilt a decision in a group of agents that employs a group decision making mechanism, we show that Really-Aware Social Choice can remain sybil-safe in the face of arbitrarily high sybil penetration, and, under certain conditions, can retain sybil-liveness.
Specifically, the problem of decision making in the presence of sybils is important in the real world. Thus, our first contribution is the development of our model which includes the status quo together with our definitions of safety, liveness, and conservatism. As such, our model allows for developing sybil-resilient rules and opens further possibilities for future study (e.g., studying ordinal elections with Borda as the base rule). Furthermore, we describe several rules and prove their sybil-resilience for important social choice scenarios; in fact, we view the simplicity of the methods as a merit, as it, e.g., allows to easily explain their operation to laymen voters.

Next we discuss pressing avenues for future research.

\mypara{Further settings}
Further research is needed to understand the possibility of sybil resilience for social choice settings other than those considered here,
such as multidimensional parameter update and multiwinner elections.
Furthermore, other types of elections (besides here we considered 1-Approval and ordinal elections) deserve study; e.g.,
cumulative voting (and also quadratic voting~\cite{lalley2018quadratic}) allows minorities to concentrate their voting power. To counter this, sybil-resilient cumulative/quadratic voting might take an approach similar to the approach for sybil-resilient parameter update, by ``suppressing $\sigma$-most lucrative voters''.

\mypara{Further base rules}
While exploring additional settings we should also explore appropriate base rules for such settings.  In addition, we should explore additional base rules for the settings at hand.

\mypara{Mitigating partial participation}
While sybils have clear incentive to vote on issues they wish to control,
genuine agents, especially if operated by humans,  might be less motivated. Thus, there is a need to augment the analysis described here to this more realistic setting.
One approach would be to use vote delegation, as is done, e.g., in liquid democracy.

\mypara{Acknowledgements}  We thank the generous support of the Braginsky Center for the Interface between Science and the Humanities.

%


\bibliographystyle{named}
\bibliography{bib}

\begin{thebibliography}{}

\bibitem[\protect\citeauthoryear{Alvisi \bgroup \em et al.\egroup
  }{2013}]{sybilsurvey}
L.~Alvisi, A.~Clement, A.~Epasto, S.~Lattanzi, and A.~Panconesi.
\newblock So{K}: The evolution of sybil defense via social networks.
\newblock In {\em Proceedings of SP '13}, pages 382--396, 2013.

\bibitem[\protect\citeauthoryear{Black}{1948}]{black1948rationale}
D.~Black.
\newblock On the rationale of group decision-making.
\newblock {\em Journal of political economy}, 56(1):23--34, 1948.

\bibitem[\protect\citeauthoryear{Bredereck \bgroup \em et al.\egroup
  }{2017}]{bredereck2017robustness}
R.~Bredereck, P.~Faliszewski, A.~Kaczmarczyk, R.~Niedermeier, P.~Skowron, and
  N.~Talmon.
\newblock Robustness among multiwinner voting rules.
\newblock In {\em Proceedings of SAGT '17}, pages 80--92, 2017.

\bibitem[\protect\citeauthoryear{Cao \bgroup \em et al.\egroup
  }{2012}]{cao2012aiding}
Q.~Cao, M.~Sirivianos, X.~Yang, and T.~Pregueiro.
\newblock Aiding the detection of fake accounts in large scale social online
  services.
\newblock In {\em Proceedings of NSDI '12}, pages 15--15, 2012.

\bibitem[\protect\citeauthoryear{Chiang \bgroup \em et al.\egroup
  }{2013}]{chiang2013secure}
J.~T. Chiang, Y.-C. Hu, and P.~Yadav.
\newblock Secure cooperative spectrum sensing based on sybil-resilient
  clustering.
\newblock In {\em Proceedings of GLOBECOM '13}, pages 1075--1081, 2013.

\bibitem[\protect\citeauthoryear{Conitzer and Yokoo}{2010}]{conitzer2010using}
V.~Conitzer and M.~Yokoo.
\newblock Using mechanism design to prevent false-name manipulations.
\newblock {\em {AI} magazine}, 31(4):65--78, 2010.

\bibitem[\protect\citeauthoryear{Conitzer \bgroup \em et al.\egroup
  }{2010}]{conitzer2010false}
V.~Conitzer, N.~Immorlica, J.~Letchford, K.~Munagala, and L.~Wagman.
\newblock False-name-proofness in social networks.
\newblock In {\em Proceedings of WINE '10}, pages 209--221, 2010.

\bibitem[\protect\citeauthoryear{Conitzer}{2008}]{conitzer2008anonymity}
V.~Conitzer.
\newblock Anonymity-proof voting rules.
\newblock In {\em Proceedings of WINE '08}, pages 295--306, 2008.

\bibitem[\protect\citeauthoryear{Douceur}{2002}]{douceur2002sybil}
J.~R. Douceur.
\newblock The sybil attack.
\newblock In {\em Proceedings of IPTPS '02}, pages 251--260, 2002.

\bibitem[\protect\citeauthoryear{Faliszewski and
  Rothe}{2016}]{controlandbribery}
P.~Faliszewski and J.~Rothe.
\newblock Control and bribery in voting.
\newblock In H.~Moulin, F.~Brandt, V.~Conitzer, U.~Endriss, A.~D. Procaccia,
  and J.~Lang, editors, {\em Handbook of Computational Social Choice}.
  Cambridge University Press, 2016.

\bibitem[\protect\citeauthoryear{Faliszewski \bgroup \em et al.\egroup
  }{2017}]{faliszewski2017bribery}
P.~Faliszewski, P.~Skowron, and N.~Talmon.
\newblock Bribery as a measure of candidate success: Complexity results for
  approval-based multiwinner rules.
\newblock In {\em Proceedings of AAMAS '17}, pages 6--14, 2017.

\bibitem[\protect\citeauthoryear{Gehrlein}{1985}]{gehrlein1985condorcet}
W.~V. Gehrlein.
\newblock The {C}ondorcet criterion and committee selection.
\newblock {\em Mathematical Social Sciences}, 10(3):199--209, 1985.

\bibitem[\protect\citeauthoryear{Lalley and Weyl}{2018}]{lalley2018quadratic}
S.~P. Lalley and E.~G. Weyl.
\newblock Quadratic voting: How mechanism design can radicalize democracy.
\newblock In {\em AEA}, volume 108, pages 33--37, 2018.

\bibitem[\protect\citeauthoryear{May}{1952}]{may1952set}
K.~O. May.
\newblock A set of independent necessary and sufficient conditions for simple
  majority decision.
\newblock {\em Econometrica}, pages 680--684, 1952.

\bibitem[\protect\citeauthoryear{McLean}{1990}]{mclean1990borda}
Iain McLean.
\newblock The {B}orda and {C}ondorcet principles: three medieval applications.
\newblock {\em Social Choice and Welfare}, 7(2):99--108, 1990.

\bibitem[\protect\citeauthoryear{Molavi~Kakhki \bgroup \em et al.\egroup
  }{2013}]{molavi2013iolaus}
A.~Molavi~Kakhki, C.~Kliman-Silver, and A.~Mislove.
\newblock Iolaus: Securing online content rating systems.
\newblock In {\em Proceedings of WWW '13}, pages 919--930, 2013.

\bibitem[\protect\citeauthoryear{Poupko \bgroup \em et al.\egroup
  }{2019}]{poupko2019sybil}
O.~Poupko, G.~Shahaf, E.~Shapiro, and N.~Talmon.
\newblock Sybil-resilient conductance-based community expansion.
\newblock In {\em Proceedings of CSR '19}, 2019.
\newblock To appear.

\bibitem[\protect\citeauthoryear{Shapiro and Talmon}{2018}]{rasc}
E.~Shapiro and N.~Talmon.
\newblock Incorporating reality into social choice.
\newblock In {\em Proceedings of AAMAS '18}, pages 1188--1192, 2018.

\bibitem[\protect\citeauthoryear{Tran \bgroup \em et al.\egroup
  }{2009}]{tran2009sybil}
D.~N. Tran, B.~Min, J.~Li, and L.~Subramanian.
\newblock Sybil-resilient online content voting.
\newblock In {\em Proceedings of NSDI '09}, pages 15--28, 2009.

\bibitem[\protect\citeauthoryear{Viswanath \bgroup \em et al.\egroup
  }{2010}]{viswanath2010analysis}
B.~Viswanath, A.~Post, K.~P. Gummadi, and A.~Mislove.
\newblock An analysis of social network-based sybil defenses.
\newblock {\em ACM SIGCOMM Computer Communication Review}, 40(4):363--374,
  2010.

\bibitem[\protect\citeauthoryear{Waggoner \bgroup \em et al.\egroup
  }{2012}]{waggoner2012evaluating}
B.~Waggoner, L.~Xia, and V.~Conitzer.
\newblock Evaluating resistance to false-name manipulations in elections.
\newblock In {\em Proceedings of AAAI '12}, pages 1485--1491, 2012.

\bibitem[\protect\citeauthoryear{Wagman and Conitzer}{2008}]{wagman2008optimal}
L.~Wagman and V.~Conitzer.
\newblock Optimal false-name-proof voting rules with costly voting.
\newblock In {\em Proceedings of AAAI '08}, pages 190--195, 2008.

\bibitem[\protect\citeauthoryear{Wagman and Conitzer}{2014}]{wagman2014false}
L.~Wagman and V.~Conitzer.
\newblock False-name-proof voting with costs over two alternatives.
\newblock {\em International Journal of Game Theory}, 43(3):599--618, 2014.

\end{thebibliography}

\end{document}